\documentclass[manuscript]{aastex631}
\usepackage{CJK}
\usepackage{natbib}
\setcitestyle{authoryear, round}

\usepackage{array} 
\usepackage{booktabs} 
\usepackage{enumitem}

\shorttitle{WFST real-time pipeline}
\shortauthors{Cai et al.}

\graphicspath{{./}{figures/}}

\begin{document}
\begin{CJK*}{UTF8}{gbsn}

\title{The 2.5-meter Wide Field Survey Telescope Real-time Data Processing Pipeline I: From raw data to alert distribution}

\correspondingauthor{Lulu Fan}
\email{llfan@ustc.edu.cn}

\author[0000-0002-0786-7307]{Minxuan Cai}
\affiliation{Department of Astronomy, University of Science and Technology of China, Hefei 230026, China}
\affiliation{School of Astronomy and Space Science, University of Science and Technology of China, Hefei 230026, China}

\author{Zelin Xu (徐则林)}
\affiliation{Department of Astronomy, University of Science and Technology of China, Hefei 230026, China}
\affiliation{School of Astronomy and Space Science, University of Science and Technology of China, Hefei 230026, China}

\author[0000-0003-4200-4432]{Lulu Fan (范璐璐)}
\affiliation{Department of Astronomy, University of Science and Technology of China, Hefei 230026, China}
\affiliation{School of Astronomy and Space Science, University of Science and Technology of China, Hefei 230026, China}
\affiliation{Deep Space Exploration Laboratory, Hefei 230088, China} 

\author[0000-0002-3105-3821]{Zhen Wan (宛振)}
\affiliation{Department of Astronomy, University of Science and Technology of China, Hefei 230026, China}
\affiliation{School of Astronomy and Space Science, University of Science and Technology of China, Hefei 230026, China}

\author{Binyang Liu (刘滨阳)}
\affiliation{Purple Mountain Observatory, Chinese Academy of Sciences, Nanjing 210023, China}

\author[0000-0002-7660-2273]{Xu Kong (孔旭)}
\affiliation{Department of Astronomy, University of Science and Technology of China, Hefei 230026, China}
\affiliation{School of Astronomy and Space Science, University of Science and Technology of China, Hefei 230026, China}
\affiliation{Deep Space Exploration Laboratory, Hefei 230088, China} 

\author[0000-0003-3424-3230]{Weida Hu (胡维达)}
\affiliation{Department of Physics and Astronomy, Texas A\&M University, College Station, TX 77843-4242, USA}
\affiliation{George P. and Cynthia Woods Mitchell Institute for Fundamental Physics and Astronomy, Texas A\&M University, College Station, TX 77843-4242, USA}

\author[0000-0001-7201-1938]{Lei Hu (胡镭)}
\affiliation{McWilliams Center for Cosmology, Department of Physics, Carnegie Mellon University, 5000 Forbes Ave, Pittsburgh, 15213, PA, USA}
\affiliation{Purple Mountain Observatory, Chinese Academy of Sciences, Nanjing 210023, China}

\author[0000-0003-0694-8946]{Qing-feng Zhu (朱青峰)}
\affiliation{Department of Astronomy, University of Science and Technology of China, Hefei 230026, China}
\affiliation{School of Astronomy and Space Science, University of Science and Technology of China, Hefei 230026, China}

\author{Guoliang Li (李国亮)}
\affiliation{Purple Mountain Observatory, Chinese Academy of Sciences, Nanjing 210023, China}

\author[0000-0003-3965-6931]{Jie Lin (林杰)}
\affiliation{Department of Astronomy, University of Science and Technology of China, Hefei 230026, China}
\affiliation{School of Astronomy and Space Science, University of Science and Technology of China, Hefei 230026, China}

\author[0000-0001-8060-1321]{Min Fang (房敏)}
\affiliation{Purple Mountain Observatory, Chinese Academy of Sciences, Nanjing 210023, China}

\author[0000-0002-9092-0593]{Ji-an Jiang (姜继安)}
\affiliation{Department of Astronomy, University of Science and Technology of China, Hefei 230026, China}
\affiliation{School of Astronomy and Space Science, University of Science and Technology of China, Hefei 230026, China}

\author[0000-0002-1935-8104]{Yongquan Xue (薛永泉)}
\affiliation{Department of Astronomy, University of Science and Technology of China, Hefei 230026, China}
\affiliation{School of Astronomy and Space Science, University of Science and Technology of China, Hefei 230026, China}

\author{Xianzhong Zhen (郑宪忠)}
\affiliation{Institute and Key Laboratory for Particle Physics, Astrophysics and Cosmology, Ministry of Education, Shanghai Jiao Tong University, Shanghai, 201210, China}

\author{Tinggui Wang (王挺贵)}
\affiliation{Department of Astronomy, University of Science and Technology of China, Hefei 230026, China}
\affiliation{School of Astronomy and Space Science, University of Science and Technology of China, Hefei 230026, China}
\affiliation{Deep Space Exploration Laboratory, Hefei 230088, China}

\begin{abstract}

The Wide Field Survey Telescope (\textsc{WFST}) is a dedicated photometric surveying facility built jointly by the University of Science and Technology of China (USTC) and the Purple Mountain Observatory (PMO). Since many of its scientific objectives rely on near-real-time data for effective analysis, prompt processing of \textsc{WFST} images is of great significance. To meet this need, we adapted the Rubin Observatory Legacy Survey of Space and Time (\textsc{LSST}) science pipelines to handle the data collected by \textsc{WFST}. This paper presents the complete data processing workflow, from ingestion of raw images to the distribution of alerts, and details the primary data products generated by our pipeline. Researchers using data processed by this pipeline can refer to this document to fully understand the data processing procedures.

\end{abstract}

\keywords{Wide-field telescopes (1800); Sky surveys (1464); Astronomy data reduction (1861); Astronomy image processing (2306); Photometry (1234) }

\section{Introduction} \label{sec:intro}

In recent decades, the field of astronomy has witnessed remarkable progress in observational capabilities. This has led to a surge in the use of various instruments for wide-field sky surveys. These instruments, as reviewed in \citet{10.1111/j.1468-4004.2007.48327.x} and \citet{Djorgovski2013}, operate across a broad spectrum from radio to gamma rays. They are designed to explore fundamental scientific questions related to the large-scale structure of the universe, the cosmic microwave background radiation, and the origin of dark matter, all of which necessitate wide-field observations.

The past two decades have seen the rise of time-domain surveys, transforming sky surveys from single-epoch wide-field observations to multi-epoch ones. This shift has expanded the scope of astronomical research from a static view of the universe to the study of the dynamic, variable sky. It allows researchers to investigate celestial objects across different time scales. For example, the Panoramic Survey Telescope and Rapid Response System \citetext{\textsc{Pan-STARRS}; \citealp{10.1117/12.552472}} uses two 1.8-meter telescopes to search for near-Earth asteroids and comets. The Palomar Transient Factory \citetext{\textsc{PTF}; \citealp{Law_2009}} and its successor, the Zwicky Transient Facility \citetext{\textsc{ZTF}; \citealp{Bellm_2019}}, have surveyed $3\pi$ of the sky at a cadence ranging from every three days to once a week. This enables the study of diverse transient phenomena, such as near-Earth asteroids and distant superluminous supernovae. The Sloan Digital Sky Survey \citetext{\textsc{SDSS}; \citealp{Gunn_2006}} has repeatedly mapped a quarter of the entire sky, enhancing our understanding of galactic and quasar physics. The Catalina Real-Time Survey \citetext{\textsc{CRTS}; \citealp{Drake_2009}} employs three wide-field telescopes to search for rare bright transients in a 33000-square-degree sky area. Currently, the most ambitious wide-field time-domain survey is the Legacy Survey of Space and Time (\textsc{LSST}) by the Vera C. Rubin Observatory \citep{LSST_science_book}. Scheduled to start operation in 2025, it features an 8-meter telescope with $u$, $g$, $r$, $i$, $z$, and $y$ filters. The \textsc{LSST} aims to achieve a single-frame $r$-band depth of about 24.5 mag, reaching approximately 27.5 mag after multiple passes. With these high-quality data, researchers plan to study the nature of dark matter and dark energy, catalog Solar System objects, map the Milky Way's structure, and investigate optical-sky transient events.

Despite these advancements, there are currently no wide-field time-domain facilities in the northern hemisphere that can achieve a depth similar to that of the \textsc{LSST}, which mainly surveys the southern hemisphere. To fill this gap, the Wide Field Survey Telescope (\textsc{WFST}) was developed. The \textsc{WFST} scans the 2$\pi$ northern sky using four filters ($ugri$), covering more than 1000 deg$^2$ per night \citep{wfstcollaboration2023sciences,Chen2024Survey}. It shares a key characteristic with the \textsc{LSST}: a wide-field survey capability.

The \textsc{LSST} has a software suite, the LSST Science Pipelines \citetext{\textsc{LSST stack}; \citealp{2017ASPC..512..279J}}, to handle its large-volume data. Due to its high portability, other observatories have adapted the \textsc{LSST stack} for their data processing. For example, the Subaru Telescope's Hyper Suprime-Cam (\textsc{HSC}) pipeline is based on the \textsc{LSST stack} \citep{HSCpipeline2018PASJ...70S...5B}, and the Gravitational Wave Optical Transient Observer (\textsc{GOTO}) has also used it for data processing \citep{GOTOpipelineii2021PASA...38...25M, GOTOpipelinei2021PASA...38....4M}. Inspired by these examples, we adopted the \textsc{LSST stack} to process \textsc{WFST} raw images.

This paper is structured as follows. Section \ref{sec:WFST_information} briefly introduces the \textsc{WFST} and its observational site conditions. Section \ref{sec:The_WFST_Data_Reduction_Pipeline} outlines the data reduction pipeline procedure and how the \textsc{LSST stack} is adapted for \textsc{WFST} data. Section \ref{sec:Steps_of_Data_Process} details the raw data processing steps. Section \ref{sec:data_products} describes the main science products generated by the pipeline. Finally, Section \ref{sec:Summary} summarizes the paper.

\section{WFST information}
\label{sec:WFST_information}

The 2.5-meter-aperture Wide Field Survey Telescope ( \textsc{WFST}) is situated at the peak of Saishiteng Mountain, near Lenghu Town in Qinghai Province, western China (93$^\circ$53$^\prime$ E, 38$^\circ$36$^\prime$ N), at an elevation of 4200 meters. The observing conditions at this location are among the most favorable globally for astronomical research, as indicated by a median seeing value of 0.75 arcseconds, an average night sky background brightness of approximately 22.0 mag arcsec$^{-2}$ in the $V$-band, and a clear sky fraction of approximately 70\% \citep{Deng2021Nature}. 

The \textsc{WFST} is designed to integrate a wide field of view with high resolution \citep{Lou2016WFSToptics,Lou2020WFST}. It encompasses a field of view of approximately 7 square degrees and is equipped with nine 85 megapixel charge-coupled devices (CCDs), each featuring a plate scale of 0.332 arcseconds per pixel \citep{Zhang2024Camera}. These CCDs are arranged in a non-contiguous configuration, resulting in two vertical and two horizontal gaps between them in each exposure. The WFST is fitted with a filter wheel that includes the $u$, $g$, $r$, $i$ and $z$ filters, along with an additional sixth $w$ filter, which is a broadband filter that spans the entire optical spectrum. The system is expected to achieve a 5$\sigma$ limiting magnitude of 23.42 mag in the $g$ band for a 30-second exposure and 25.95 mag for a 50-minute stacked exposure \citep{Lei_2023}. 

Capitalizing on its extensive field of view and high observation cadence, the WFST's primary scientific objectives in time-domain science encompass the investigation of supernovae \citep{2022Univ....9....7H}, tidal disruption events \citep{2022MNRAS.513.2422L}, multi-messenger events \citep{ Liu2023KN, Yu2024GW}, variable stars \citep{Lin2024}, and active galactic nuclei \citep{Su2024AGN}. Furthermore, the WFST will make significant contributions to research on asteroids and other bodies within the Solar System \citetext{Wang et al. in prep, \citealp{ljq2025}}, the global cluster (Wan et al. in prep), the structural composition of the Milky Way and its dwarf satellite galaxies, galaxy formation, and cosmology \citep{wfstcollaboration2023sciences}.

\section{The WFST Data Reduction Pipeline}
\label{sec:The_WFST_Data_Reduction_Pipeline}

\begin{figure*}
    \centering
    \includegraphics[width=0.8\linewidth]{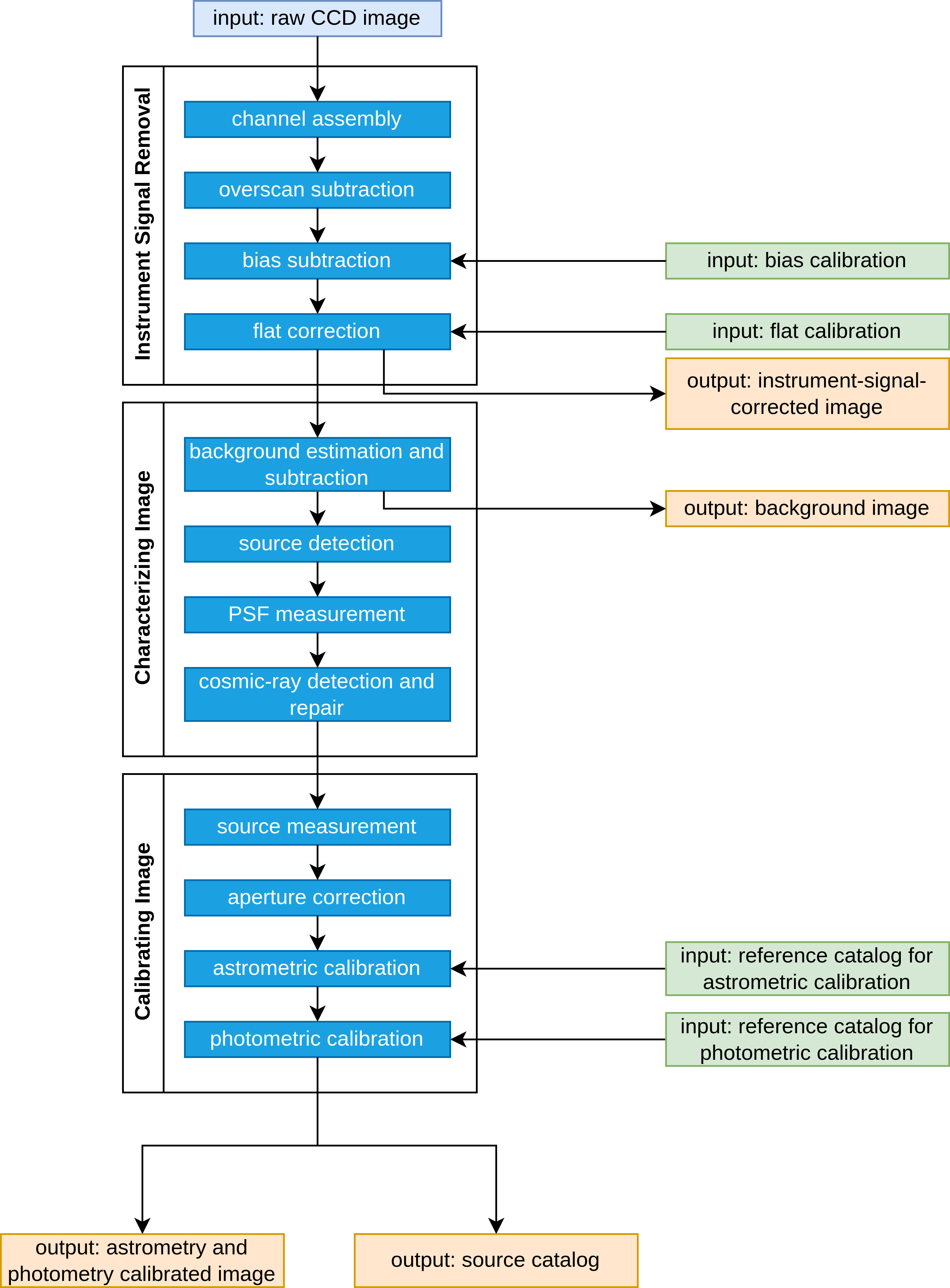}
    \caption{Processing flow of the single-frame-process phase}
    \label{fig:pipeline_intro_single_frame_flow}
\end{figure*}

\begin{figure*}
    \centering
    \includegraphics[width=0.8\linewidth]{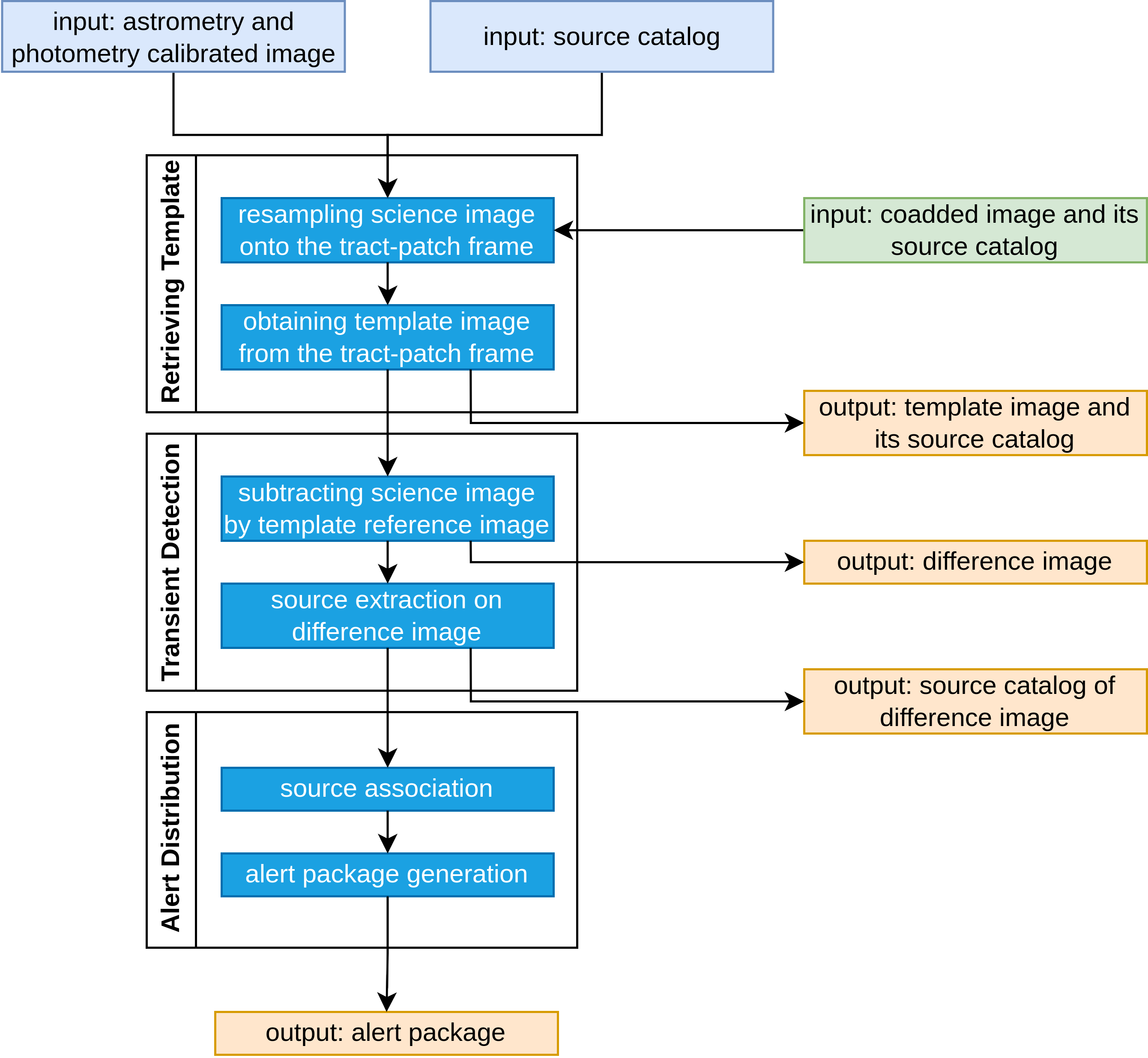}
    \caption{Processing flow of the image-differencing phase}
    \label{fig:pipeline_intro_difference_flow}
\end{figure*}

Figures \ref{fig:pipeline_intro_single_frame_flow} and \ref{fig:pipeline_intro_difference_flow} represent the comprehensive methodology of our WFST data reduction pipeline. Upon acquisition of the raw CCD image files from the Lenghu observatory site, this pipeline commences the processing sequence. Subsequent to the removal of instrument signals, the pipeline undertakes image characterization, calibration, transient detection, and alert distribution, culminating in the generation of a suite of scientific products (see Section \ref{sec:data_products}). These products are subsequently transferred expeditiously to the data server. We have selected the  \textsc{LSST} stack for the implementation of these data reduction procedures, given its particular suitability for the outlined processes. In the ensuing section, we will provide an overview of the \textsc{LSST} stack and delineate its application in the data processing operations of \textsc{WFST} raw images.

\subsection{The \textsc{LSST} Stack and the {\texttt {OBS\_WFST}} Package}

The \textsc{LSST} stack capitalizes on efficiency and adaptability through the use of foundational algorithms implemented in \verb!C++! and higher-level tasks realized in Python. The architecture ensures high portability, enabling users to concentrate on integrating top-level Python tasks without the necessity of comprehending \verb!C++!. 

When employing the algorithms within the \textsc{LSST} stack to process imagery from distinct telescopes, users are only required to develop an `obs package' tailored to the specific attributes of their telescopes. In our instance, this package is designated as \texttt{obs\_wfst} and employs the `Generation 3' Butler for data organization and retrieval. The \texttt{obs\_wfst} package comprises several key components, including a set of configuration files to define processing steps, a set of pipeline files to organize task execution sequences, Python scripts for adjusting or revising standard processes, and a description of detectors. 

The configuration files, preserved in Python script format, contain settings that are derived from their corresponding \textsc{LSST}-provided tasks. These configuration files supersede the original task configurations, permitting us to adjust pipeline parameters directly within these configuration files, as opposed to modifying the foundational task files. For instance, to establish a threshold value of 5 for source detection, it is sufficient to assign \texttt{config.detection.thresholdValue = 5.0} within the pertinent configuration file. 

While the configuration files dictate task parameters, the pipeline files arrange the procedures of the tasks. These are stored in YAML format, enabling users to configure subtasks offered by the \textsc{LSST} stack (or their own custom tasks) into the main task they require. For example, our \texttt{processCcd} task incorporates the \texttt{WfcIsrTask} (which inherits from \texttt{IsrTask}), \texttt{CharacterizeImageTask}, and \texttt{CalibrateTask}, with these subtasks being automatically interconnected through their input/output relations when the pipeline is executed with the \texttt{processCcd} task. 

Although modifications to the foundational tasks provided by the \textsc{LSST} stack may be necessary, users have the option to author Python scripts to customize these tasks by deriving from the class-based tasks provided by the stack or even to create new tasks altogether. In our scenario, we have either inherited tasks (such as \texttt{WfcIsrTask} deriving from \texttt{IsrTask}, and \texttt{WfcDiaPipelineTask} deriving from \texttt{DiaPipelineTask}) or devised new tasks (such as \texttt{SFFTSubtractTask}, which executes image subtraction using an innovative method detailed in Section \ref{sec:Subtract_Image_and_Transient_Detection}, a method not afforded by the \textsc{LSST} stack). This methodology increases the versatility of the \textsc{LSST} stack for processing our \textsc{WFST} data. 

Furthermore, users are required to prepare description files for each detector in the Flexible Image Transport System (FITS) format to ensure that the \textsc{LSST} stack accurately understands the properties of the novel camera. To enhance processing efficiency, specifically concerning the \textsc{WFST}, each CCD (consisting of $9216 \times 9232$ pixels) is partitioned into four smaller sub-CCDs, each measuring $4608 \times 4616$ pixels. These sub-CCDs, herein referred to as CCD-quadrants, facilitate parallel processing and improve overall efficiency. Consequently, we delineate 36 detector files, with every four detectors corresponding to a single CCD. All subsequent data products (see a detailed description in Section \ref{sec:data_products}) are produced on a per-CCD quadrant basis.

\section{Steps of Data Processing}
\label{sec:Steps_of_Data_Process}

Upon capturing each exposure with the WFST wide field camera (WFC) at Lenghu observatory site, the corresponding data is archived in FITS format within a span of 10 seconds post shutter closure. Subsequently, the file undergoes compression and is immediately transmitted to the data reduction server situated in the data center of USTC. Once received and decompressed, these data are conveyed to the data reduction pipeline for near-real-time processing. In this section, we delineate the procedures involved in managing the WFST data through the \textsc{LSST} stack, encompassing the process from the ingestion of raw images to the production of the final data output (see Figures \ref{fig:pipeline_intro_single_frame_flow} and \ref{fig:pipeline_intro_difference_flow} for further details).

\subsection{Raw Image Ingestion}
\label{sec:raw_image_ingestion}

Upon reception of an exposure, the data reduction pipeline initiates the creation of a temporary workspace for processing the raw image. This workspace constitutes a pre-configured \textsc{LSST} stack environment, encompassing essential elements such as the master bias and flat calibration images (see Section \ref{sec:construction_of_calibration_frames}), coadd images, source catalogs, and reference catalogs necessary for processing the \textsc{WFST} data from that observational night. This workspace is strategically prepared in advance, ahead of the observational night. 

Following the establishment of the temporary workspace, the process of raw image ingestion is executed. This procedure entails extracting pertinent information from the headers of raw FITS files, encompassing observation type, band, visit number, CCD number, exposure time, azimuth, altitude, right ascension, and declination, and recording it within a \texttt{sqlite3} database. Concurrently, these files are linked from their original storage locations to the workspace, in readiness for processing by the \textsc{LSST} stack.

\subsection{Construction of Calibration Frames}
\label{sec:construction_of_calibration_frames}

For each observational night, bias and flat frames are subjected to a median combination and subsequently incorporated into the processing pipeline. Initially, a designated workspace is established for the assembly of calibration frames. The raw bias and flat images are then ingested into this workspace (see Section \ref{sec:raw_image_ingestion}). Depending on the type of observation recorded in the \texttt{sqlite3} database within the workspace, the \textsc{LSST} stack discerns and processes the bias and flat frames independently. 

During the calibration procedure, the raw bias images undergo overscan subtraction, followed by a median stacking to produce the master bias image. Consequently, employing the master bias image, raw flat images are processed through overscan and bias subtraction, culminating in median stacking to produce the master flat image. Once the master calibration frames are constructed, they are assimilated into the provisional workspace delineated in Section \ref{sec:raw_image_ingestion}. These calibration products are subsequently used to process the raw science images.

\subsection{Single Frame Processing}
\label{sec:single_frame_process}

Following the construction and ingestion of the calibration frames, the raw images can be processed in the temporary workspace using the \textsc{LSST} stack, starting with single-frame processing. This phase entails the removal of instrumental signals by subtracting the overscan and implementing the master bias and flat corrections to mitigate instrumental artifacts. Subsequently, the images undergo characterization through background estimation and subtraction, source detection, point source function (PSF) measurement, and cosmic ray detection and elimination. Ultimately, images are calibrated by performing source measurements, applying aperture corrections, and performing both astrometric and photometric calibrations. These steps are detailed in the following subsections.

\subsubsection{Instrumental Signal Removal}
\label{sec:instrumental_signal_removal}

At this stage, the science frames undergo correction using the constructed master bias and flat frames (see Section \ref{sec:construction_of_calibration_frames}). Initially, each CCD is partitioned into four quadrants via channel assembly, concurrently facilitating overscan subtraction. Subsequently, bias subtraction is executed, followed by normalization with the master flat images to address pixel response non-uniformities. For the processes of bias subtraction and flat correction, we maintained the default parameters because the \textsc{LSST} stack, under these conditions, yielded satisfactory outcomes.

Given the limitation of the default instrumental signal removal task in executing the overscan subtraction both horizontally and vertically, we have developed \texttt{WfcIsrTask}, an advancement derived from \texttt{IsrTask}, specifically designed to meet this requirement. The enhanced task gives the ability to subtract the overscan in dual directions, which can be controlled through the newly introduced \texttt{doBothOverscan} option. Furthermore, the peripheral corners of the four CCDs at the margins of the \textsc{WFST} field of view exceed the usable field boundaries. To address this issue, these regions are masked during the instrumental signal removal process, a feature that can be activated or deactivated using the \texttt{doCorners} option.

\subsubsection{Characterizing Image}
\label{sec:characterize_iamge}

In this stage, the background is quantified and subsequently subtracted. During the source detection process, the PSF is constructed and cosmic rays are identified and eliminated. The \textsc{LSST} stack employs sixth-order Chebyshev polynomials to model the background in both the X and the Y directions. During the modeling process, each CCD quadrant is partitioned into regions measuring 512 by 512 pixels, a configuration that produces satisfactory results for the images. Subsequent to the instrumental signal removal, the modeled background is subtracted from the image.

Following background subtraction, the \textsc{LSST} stack evaluates the PSF information across the image. The PSF is determined by fitting the point sources identified within the image. An initial detection of these point sources is performed using a detection threshold established at 5$\sigma$ and 4 pixels. Under optimal weather conditions, this threshold typically guarantees a sufficient number of sources (exceeding 2) required for precise PSF fitting. The \texttt{psfex} algorithm (\texttt{lsst.meas.extensions.psfex.psfexPsfDeterminer}) is utilized during the PSF measurement phase, applying its default parameters. Subsequent to PSF modeling, the detection and elimination of cosmic rays are conducted. Pixels contaminated by cosmic rays are identified and subjected to interpolation. Thereafter, these interpolated pixels are flagged as ``interpolated".

\subsubsection{Calibrating Image}
\label{sec:calibrate_image}

Subsequent to the processes of background subtraction, PSF model construction, and cosmic-ray mitigation, the single frame is subjected to procedures of source measurement, aperture correction, as well as astrometric and photometric calibration. This phase guarantees the precision of the World Coordinate System (WCS) information and determines the magnitude zero point for each CCD-quadrant. 

During the source measurement stage, a variety of algorithms are employed, which encompass PSF model fitting, Kron photometry, and aperture photometry utilizing radii of 3.0, 4.5, 6.0, 9.0, 12.0, 17.0, 25.0, 35.0, 50.0, and 70.0 pixels (these corresponding to 0.996, 1.494, 1.992, 2.988, 3.984, 5.644, 8.300, 11.620, 16.600, and 23.240 arcseconds, respectively). An aperture correction is then applied to enhance the accuracy of the aperture photometry results.

The astrometric calibration process involves the correction of the WCS information (i.e., \texttt{RA}, \texttt{DEC}, and \texttt{INST-PA}) derived from the FITS header of the raw images. This correction is necessitated by minor discrepancies between the actual WCS values and those specified in the header, which arise due to temporal delays between the Operator Control System (OCS) and the Telescope Control System (TCS). Moreover, optical systems, particularly those featuring wide fields of view, are prone to image distortion. Consequently, the \textsc{LSST} stack measures these distortions during the astrometric calibration phase. This calibration is achieved by aligning the detected sources with an external reference catalog. The Gaia-DR3 catalog \citep{GAIADR3} is used for \textsc{WFST} due to its exceptional positional accuracy. The resultant data are represented in the TAN-SIP WCS format, which utilizes the Simple Imaging Polynomial (SIP) model to characterize image distortions.

Subsequent to astrometric calibration, the \textsc{LSST} stack performs photometric calibration by comparing the sources identified in previous steps with a photometric reference catalog. For the $g$, $r$, $i$, and $z$ bands, the Pan-STARRS DR1 catalog \citep{PANSTARR-PS1} is used, given its coverage of these bands. However, since the Pan-STARRS catalog does not encompass a $u$ band, the SDSS $u$-band catalog \citep{SDSSDR7} is used for the calibration of the $u$-band image. To address the discrepancies between the \textsc{WFST} and the reference catalog filters, the \texttt{photoCal.colorterms} parameter is utilized, which rectifies these differences by applying a second-order polynomial to each band. Following the photometric calibration, the determination of the magnitude zero point is performed for each CCD-quadrant, thereby establishing the correlation between instrumental flux and magnitude.

\subsection{Image Coadding}
\label{sec:image_coadding}

Following the image characterization and calibration detailed in the preceding subsections, individual science exposures are obtained. To effectively identify variable sources, these exposures are later be employed in image differencing, which produces a difference image containing variation information. Prior to conducting image differencing, it is essential to stack the science exposures to generate coadded images. 

The \textsc{LSST} stack organizes the sky into several regions referred to as tracts, each of which is subdivided into smaller regions known as patches. For the WFST, the internal dimensions of the patches are configured to be 5000 $\times$ 5000 pixels, each pixel encompassing 0.33 arcseconds. To ensure sufficient overlap, a patch border of 200 pixels and a tract overlap of 0.25$^\circ$ are defined. The patches are not rotated, and a TAN projection (the gnomonic projection, see \citealp{Calabretta+2002}) is employed. 

Before image stacking, a preliminary task, referred to herein as \texttt{selectGoodSeeingVisits}, is undertaken to select exposures with favorable seeing conditions, thereby ensuring the quality of the final coadded images. During the stacking process, each science frame is segmented according to the tract and patch configuration. The segmented components are aligned with their designated sky regions on the basis of their tract and patch assignments. The \texttt{MEANCLIP} statistic is utilized to aggregate data across epochs. After stacking, the \textsc{LSST} stack generates and archives images corresponding to each sky region in the form of tracts and patches. Following the completion of the image stacking process, source detection is conducted on the images for each patch, resulting in source catalogs for the stacked images. These source catalogs are subsequently used during the image subtraction process.

\subsection{Image Differencing}
\label{sec:image_difference}

\subsubsection{Retrieving Template}
\label{sec:retrieve_template}

Before image subtraction is performed, a template image is derived from the co-added images to correspond precisely to the dimensions and spatial positioning of the associated science image. This is achieved by employing the WCS information from the science image to delineate the specific area within the coadded image that corresponds to the science image's size and location. It should be noted that a science image may extend across the boundaries of multiple patches; in such instances, the pertinent data are sourced from these patches and concatenated to form a cohesive template image. Similarly, the source catalogs corresponding to the coadded images are extracted and amalgamated to construct the template source catalog.

\subsubsection{Subtracting Image and Transient Detection}
\label{sec:Subtract_Image_and_Transient_Detection}

The Saccadic Fast Fourier Transform (\textsl{SFFT}) algorithm \citep{2022ApJ...936..157H} is employed to subtract template images from science images. Compared to the most widely utilized AL algorithms \citep{AL_algorithm} in traditional applications, such as \textsl{HOTPANTS} software \citep{HOTPANTS_software}, the \textsl{SFFT} algorithm exhibits significant advantages primarily due to its increased accuracy and accelerated processing speed. Using a delta function convolution kernel as its basis function, \textsl{SFFT} demonstrates greater adaptability to the PSF of actual observational data relative to conventional Gaussian function bases, thus achieving more precise PSF alignment. In the context of transient source detection in the presence of  host galaxies, \textsl{SFFT} effectively mitigates the influence of residual patterns originating from the host galaxy in difference images, consequently enhancing the detection sensitivity and photometric precision for transient sources proximal to the host galaxy. The GPU-implemented algorithm \textsl{SFFT} facilitates image subtraction at a rate approximately an order of magnitude faster than the CPU-based alternatives such as \textsl{HOTPANTS} and \textsl{ZOGY} \citep{ZOGY_algorithm}, thus rendering it more suitable for meeting the real-time data processing demands of \textsc{WFST}. In practice, we employ the DCU (Deep-learning Computing Unit) K100-AI 64GB acceleration card to perform image differencing using the \textsl{SFFT}  algorithm. For our image differencing tasks, this card demonstrates performance comparable to that of the NVIDIA A100 40GB. Specifically, it achieves an average processing time of approximately 4 seconds per CCD-quadrant. Furthermore, the \textsl{SFFT} algorithm demonstrates superior versatility compared to the \textsl{HOTPANTS} algorithm, as it is applicable to both sparse and dense stellar fields, thereby better supporting the broad scientific goals of the \textsc{WFST}. It should be noted that the \textsc{LSST} stack in Generation 3 only incorporates the \textsl{HOTPANTS} algorithm, as the \textsl{ZOGY} algorithm is not supported in Version 24.0. Consequently, the \textsl{HOTPANTS} algorithm can be utilized for image subtraction in scenarios where GPU resources are not accessible. 

Thereafter, source detection is conducted on the difference images, and the data pertaining to the detected sources is systematically documented in the difference source catalog. Besides the parameters assessed in the scientific images, the difference source catalog will also encompass details regarding dipole and trail morphology, which can assist in more effectively differentiating between real and bogus sources.

\subsubsection{Source Association and Alert Distribution}
\label{sec:source_associtaion_and_alert_distribution}

Following the generation of the source catalog for the difference images, the  \textsc{LSST} stack archives pertinent source information within a database dedicated to the association of the source. Additionally, it correlates existing observational data pertaining to these sources, thereby enabling the construction of light curves for users. For the purpose of source association, a PostgreSQL database is used, which is shown to be more efficient in managing substantial data volumes in comparison to the default lightweight SQLite3 database. The results of the source association are presented in the Avro file format, with a detailed exposition provided in Section \ref{sec:data_products}.

\section{Data Products}
\label{sec:data_products}

Table \ref{tab:data_products} enumerates the main data products of our \textsc{WFST} pipeline, including details on the format and generation frequency for each product. This table also identifies the sections that outline the pertinent pipeline procedures responsible for the creation of each product. Apart from raw images, source association databases, coadded images, and coadded source catalogs, all other products are produced on an individual CCD-quadrant basis. Subsequently, we offer a comprehensive elucidation of each data product enumerated in Table \ref{tab:data_products}.

\begin{table*}[htbp] 
    \centering 
    \caption{Data Products}
    \label{tab:data_products} 
    \begin{tabular}{llcll}
        \toprule
        \toprule 
        ProductID & Description & Pipeline & Format & Generation Frequency \\
        \midrule
        1 & Raw images & Section \ref{sec:raw_image_ingestion} & FITS & Real time \\
        2 & Science images & Section \ref{sec:single_frame_process} & FITS & Real time \\
        3 & Science source catalogs & Section \ref{sec:single_frame_process} & FITS & Real time \\
        4 & Difference images & Section \ref{sec:Subtract_Image_and_Transient_Detection} & FITS & Real time \\
        5 & Difference source catalogs & Section \ref{sec:Subtract_Image_and_Transient_Detection} & FITS & Real time \\
        6 & Template images & Section \ref{sec:retrieve_template} & FITS & Real time \\
        7 & Alert packages & Section \ref{sec:source_associtaion_and_alert_distribution} & Avro & Real time \\
        8 & Source association database & Section \ref{sec:source_associtaion_and_alert_distribution} & PostgreSQL & Real time \\
        9 & Calibration images & Section \ref{sec:construction_of_calibration_frames} & FITS & Daily \\
        10 & Coadded images & Section \ref{sec:image_coadding} & FITS & Static \\
        11 & Coadded source catalogs & Section \ref{sec:image_coadding} & FITS & Static \\
        \bottomrule
    \end{tabular}
\end{table*}

\begin{enumerate}[start=1] 
    \item \textit{Raw images}. For \textsc{WFST}, each exposure produces 9 separate CCD data files. Each CCD is further divided into 16 amplifier channels, with each channel containing overscan strips. Each channel is stored in a separate FITS extension, with each extension possessing its own header. 
    \item \textit{Science images}. These are calibrated CCD-quadrant images, with astrometric and photometric calibration results included in the header. In addition, the file contains a mask image that marks bad pixels, such as saturated, edge, and interpolated pixels, a variance image representing the errors, and the results of PSF fitting.
    \item \textit{Science source catalogs}. The science source catalog contains detailed information for each source detected in the science images. This includes their right ascension and declination, pixel coordinates, and morphological parameters. It also includes instrumental fluxes measured with different aperture sizes, fluxes derived from PSF model fitting, and Kron photometry (see Section \ref{sec:calibrate_image}). Additionally, the catalog contains classification results that distinguish point sources from extended sources.
    \item \textit{Difference images}. The difference image is the result of subtracting the template image from the science image. It primarily stores information that is consistent with the science image, highlighting the variations between the science image and the template image, such as transient or variable sources.
    \item \textit{Difference source catalogs}. The difference source catalog contains the same information as the science source catalog, including details such as source coordinates, flux measurements, and morphological parameters. Additionally, the difference source catalog includes measurement results for each source using the dipole and trail models, which help to further characterize the sources and distinguish between real and bogus detections. 
    \item \textit{Template images}. These images are extracted from the coadded images using the WCS information from the corresponding science image. As a result, each template image is precisely aligned with its corresponding science image, pixel by pixel.
    \item \textit{Alert packages}. The alert distribution files of the \textsc{LSST} stack are stored in Avro format and contain detection results for each variable source. These results primarily include WCS and pixel coordinates, observation time, band, total flux, differential flux, and measurements of dipole and trail. Additionally, the files include 30 × 30 pixel image cutouts centered on the alert position from each image. The detection results cover both the current detection and previous detections of the source within the past six months.
    \item \textit{Source association database}. We use a PostgreSQL database to store information about transients generated by the \textsc{LSST} stack. This database plays a crucial role in transient association within the LSST stack and is not intended for future public release.
    \item \textit{Calibration images}. These include the master bias and master flat images, which are constructed nightly by the \textsc{LSST} stack. On a nightly basis, we generate the corresponding calibration images.
    
    \item \textit{Coadded images}. These images are stacked by the \textsc{LSST} stack. Unlike the science and difference images, the coadded images are stored on a tract/patch basis rather than a CCD-quadrant basis.
    
    \item \textit{Coadded source catalogs}. The source catalogs for coadded imagess are organized on a tract/patch basis, analogous to the configuration of the coadded images themselves.
\end{enumerate}

\section{Summary}
\label{sec:Summary}

In this paper, we have presented a concise summary of the \textsc{WFST}'s observing conditions and instrument details. Given the similarities between the WFST and the LSST projects, such as their wide fields of view, high cadences, and overlapping scientific goals, we customized the \textsc{LSST} stack through our \texttt{obs\_wfst} package to process \textsc{WFST} data. This customization allows us to effectively handle the data collected by the \textsc{WFST}. We then delved into the step-by-step process of handling each raw image, starting from ingestion, going through calibration, image subtraction, and ending with alert distribution. Each stage of the data processing pipeline was carefully described, highlighting the key algorithms and tasks involved. Furthermore, we provided an in-depth look at the main data products generated by our data reduction pipeline. These products, which include raw images, science images, source catalogs, difference images, and alert packages, among others, are essential for scientific research using \textsc{WFST} data.

In general, this paper serves as a valuable resource for researchers using data from our \textsc{WFST} data reduction pipeline. It offers detailed information on the data processing steps, allowing them to better understand and analyze the \textsc{WFST} data for their specific scientific inquiries.

\begin{acknowledgments}
The Wide Field Survey Telescope (WFST) is a joint facility of the University of Science and Technology of China, Purple Mountain Observatory. This work is supported by National Key Research and Development Program of China (2023YFA1608100). The authors appreciate the support of the National Natural Science Foundation of China (NSFC, Grant Nos. 12233008, 12173037, 12025303, 12393814), the CAS Project for Young Scientists in Basic Research (No. YSBR-092), the Fundamental Research Funds for the Central Universities (WK3440000006) and Cyrus Chun Ying Tang Foundations.
\end{acknowledgments}

\vspace{5mm}
\facilities{\textsc{WFST},\textsc{LSST}}

\software{LSST pipeline \citep{2017ASPC..512..279J}}




\bibliography{ms}{}
\bibliographystyle{aasjournal}



\end{CJK*}
\end{document}